\documentclass[final]{aipproc}
\usepackage{epsfig}
\usepackage{amssymb}
\layoutstyle{6x9}

\begin{document}

\title[What do we learn from PHENIX results on non-identified jet
correlation?]{From mach cone to reappeared jet: What do we learn
from PHENIX results on non-identified jet correlation?}

 \keywords{}

\author{Jiangyong Jia for the PHENIX Collaboration}{
  address={Columbia University, New York, NY 10027 and Nevis Laboratories, Irvington, NY 10533, USA}
}
\begin{abstract}
High $p_T$ jets are known to be strongly modified by the dense,
strongly interacting medium created in heavy-ion collisions. The
jet signal, extracted from two particle $\Delta\phi$ correlation,
shows a systematic evolution of these modifications as function of
$p_T$ and centrality. At intermediate $p_T$, both near side and
away side correlations are modified. But the modifications are
much stronger at the away side, resulting in a characteristic cone
type of structure in central Au + Au collisions. The robustness of
cone structure is strengthened by studying the jet shape as
function of angle relative to the reaction plane. As one increase
the $p_T$ for BOTH hadrons, the cone structure seems to be filled
up, and a peak structure appears on the away side. However, the
interpretation of these results require careful separation of
medium effect and surface bias.
\end{abstract}

\maketitle


\section{Introduction}
High $p_T$ back-to-back jets are valuable probes for the
sQGP~\cite{RHIC} created in heavy-ion collisions at RHIC. Existing
two particle jet correlation results from statistically limited
RUN2 Au + Au data set revealed a strong interaction of the jets
with the medium. On the one hand, jet correlation at high $p_T$
indicates a seemly complete disappearance of the away side jet
signal~\cite{Adler:2002tq}. On the other hand, jet correlation at
low $p_T$ shows an enhancement of the away side jet
yield~\cite{Adams:2005ph} but a broadened jet
shape~\cite{Adler:2005ee}. Qualitatively, this is consistent with
the energy loss picture, where the high $p_T$ jets are quenched by
the medium and their lost energy enhanced the jet multiplicity at
low $p_T$.

Equipped with excellent statistics from RUN4 Au + Au and RUN5 Cu +
Cu data sets, we would like to gain further understanding on the
interaction of the jets with the medium. We hope to address
important questions such as: How the jet looses it's energy? How
the lost energy get redistributed? How the medium responds to the
jet? What happens to the higher $p_T$ jets? We attempt to address
these questions using the non-identified charged hadron - charged
hadron correlation results from RUN4 Au + Au data set.

\section{Jet properties at intermediate $p_T$}
Our analysis is based on 1 billion minimum bias events from Au +
Au collisions at $\sqrt{s}$ = 200 GeV. The correlation function
$C(\Delta\phi)$ (CF) is defined as the ratio of same event pair
distribution, $dN^{\rm{pairs}}/d\Delta\phi$ to the mixed event
pair distribution, $dN^{\rm{mix}}/d\Delta\phi$.
$dN^{\rm{mix}}/d\Delta\phi$ reflects the level of combinatoric
background and the geometrical acceptance~\cite{Jia:2004sw}. In
heavy-ion collisions, the CF can be expressed as the sum of jets
and elliptic flow,
\begin{eqnarray}
\label{eq:1} C(\Delta\phi) = J(\Delta\phi) + \xi\left(1+2v_2^{t}
v_2^{a} \cos2\Delta \phi\right)
\end{eqnarray}
The superscript $t$ and $a$ stand for the trigger and associated
particles, $\xi$ is a normalization factor.
\begin{figure}[ht]
\epsfig{file=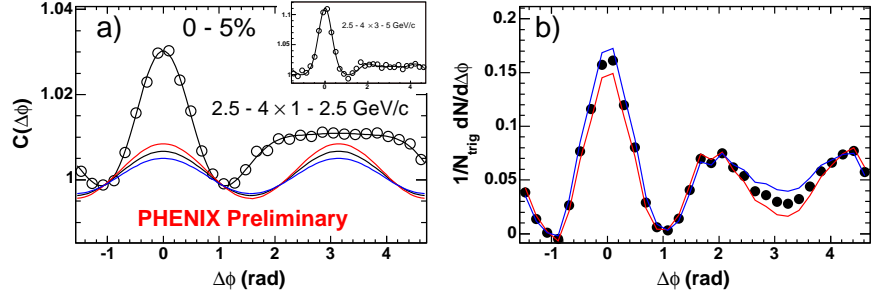,width=0.8\linewidth}
\caption{\label{fig:zyamcf1} a) Correlation function in 0-5\%
centrality bin, the lines indicated the level of flow background
and it's systematic error band, the insert is CF for higher
$p_{T,{\rm{assoc}}}$. b) Correspondingly background subtracted
per-trigger yield.}
\end{figure}

Fig.\ref{fig:zyamcf1}a shows the typical correlation function from
central Au + Au collisions. The away side shape is very broad and
non-gauss like. It has a plateau that expands to about 2 radians
and a possible small dip at $\pi$. The subtraction of the flow
contribution (shown by the curves) only makes the dip deeper
(Fig\ref{fig:zyamcf1}b). $\xi$ is fixed by scaling the flow term
to match the CF, i.e. assuming $J = 0$ at some $\Delta\phi$ (ZYAM
assumption)~\cite{Ajitanand:2005jj}. The ZYAM procedure leads to a
slight over-subtraction of jet yield, however since $2v_2^tv_2^a
\approx$ few $\%$, the over-subtraction mainly results in a
vertical shift and does not affect the away side jet shape. The
systematic error on $J(\Delta\phi)$ is dominated by the
uncertainties on $v_2$.

\begin{figure}[b]
\epsfig{file=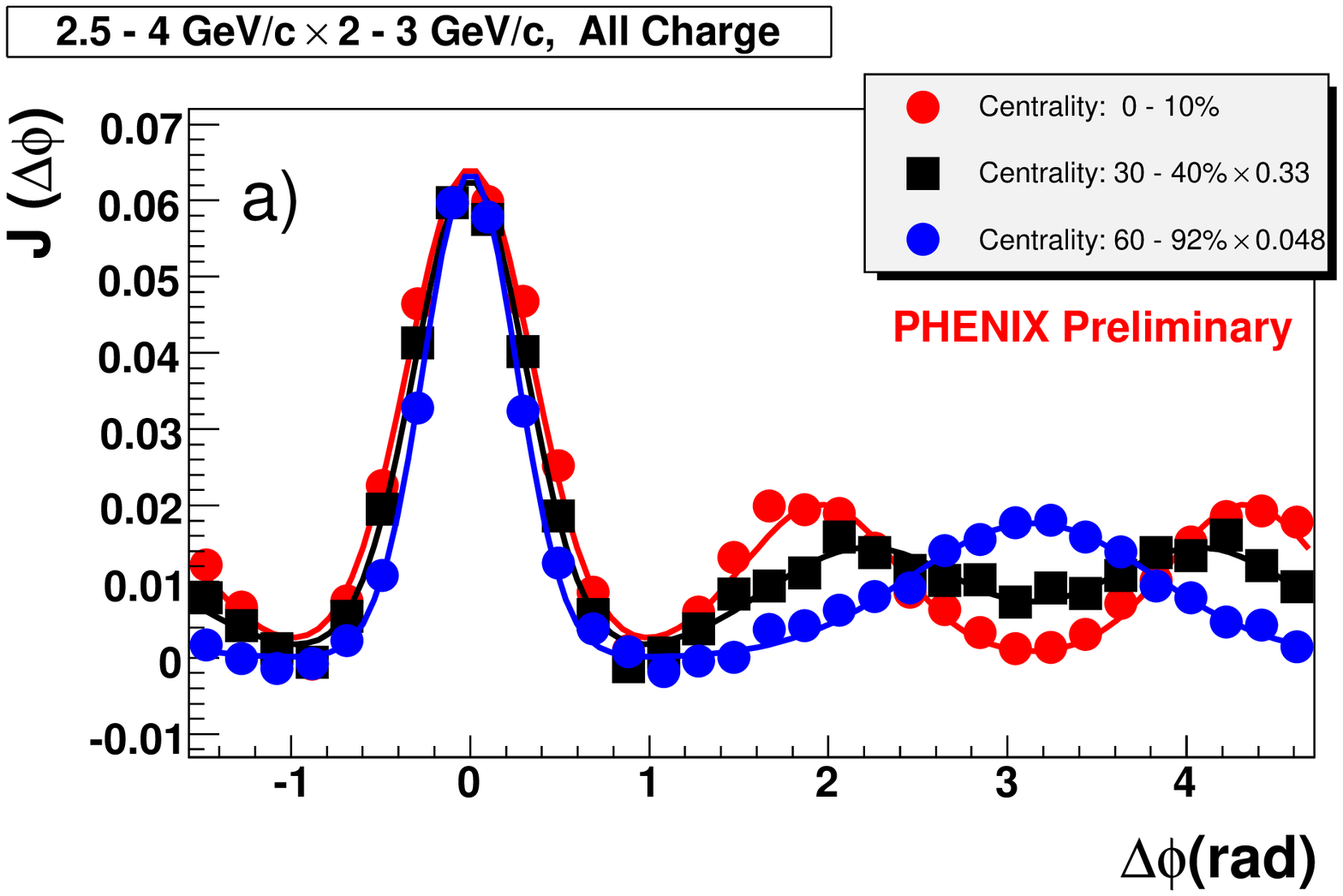,width=0.33\linewidth}
\epsfig{file=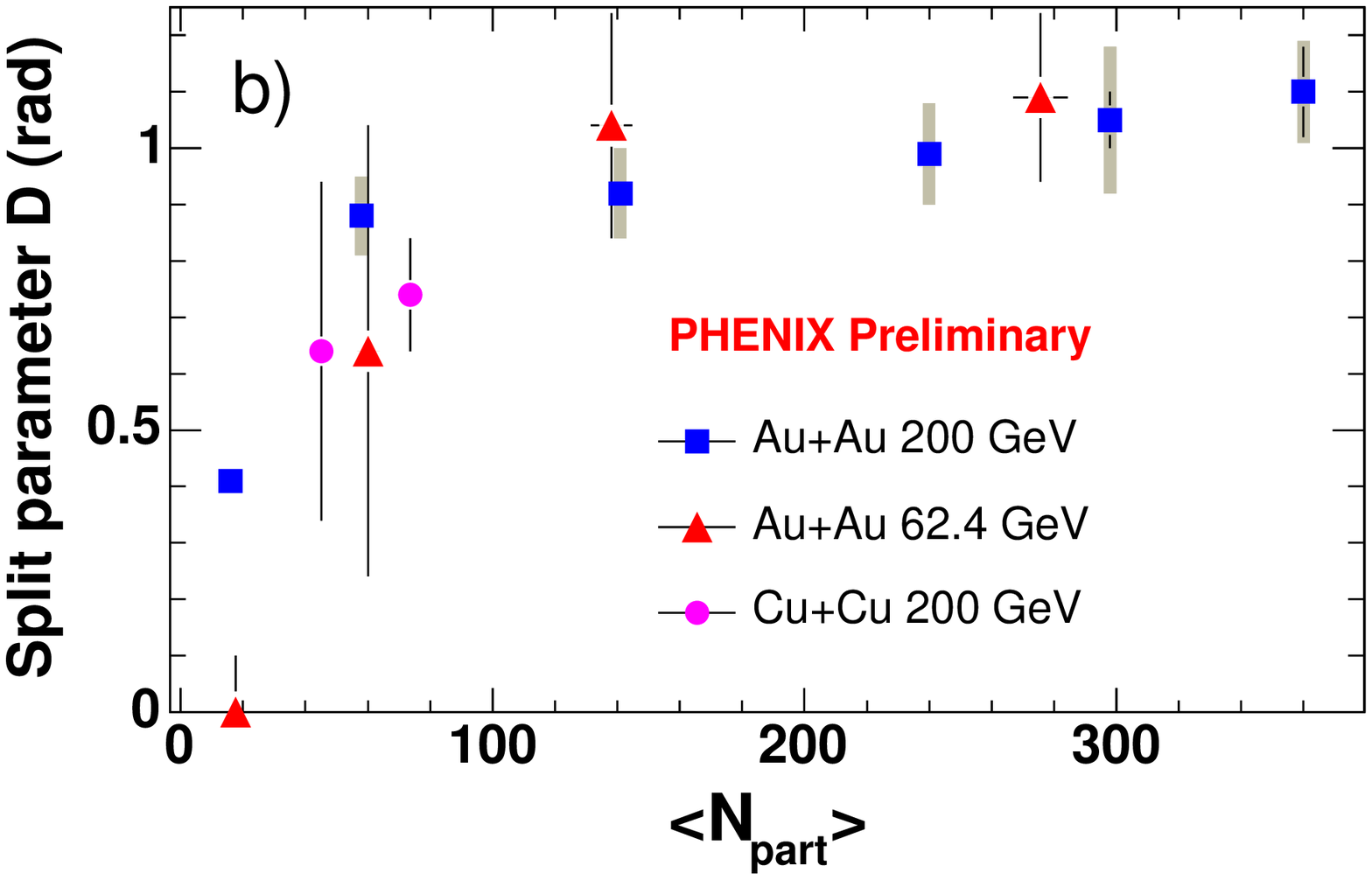,width=0.33\linewidth}
\epsfig{file=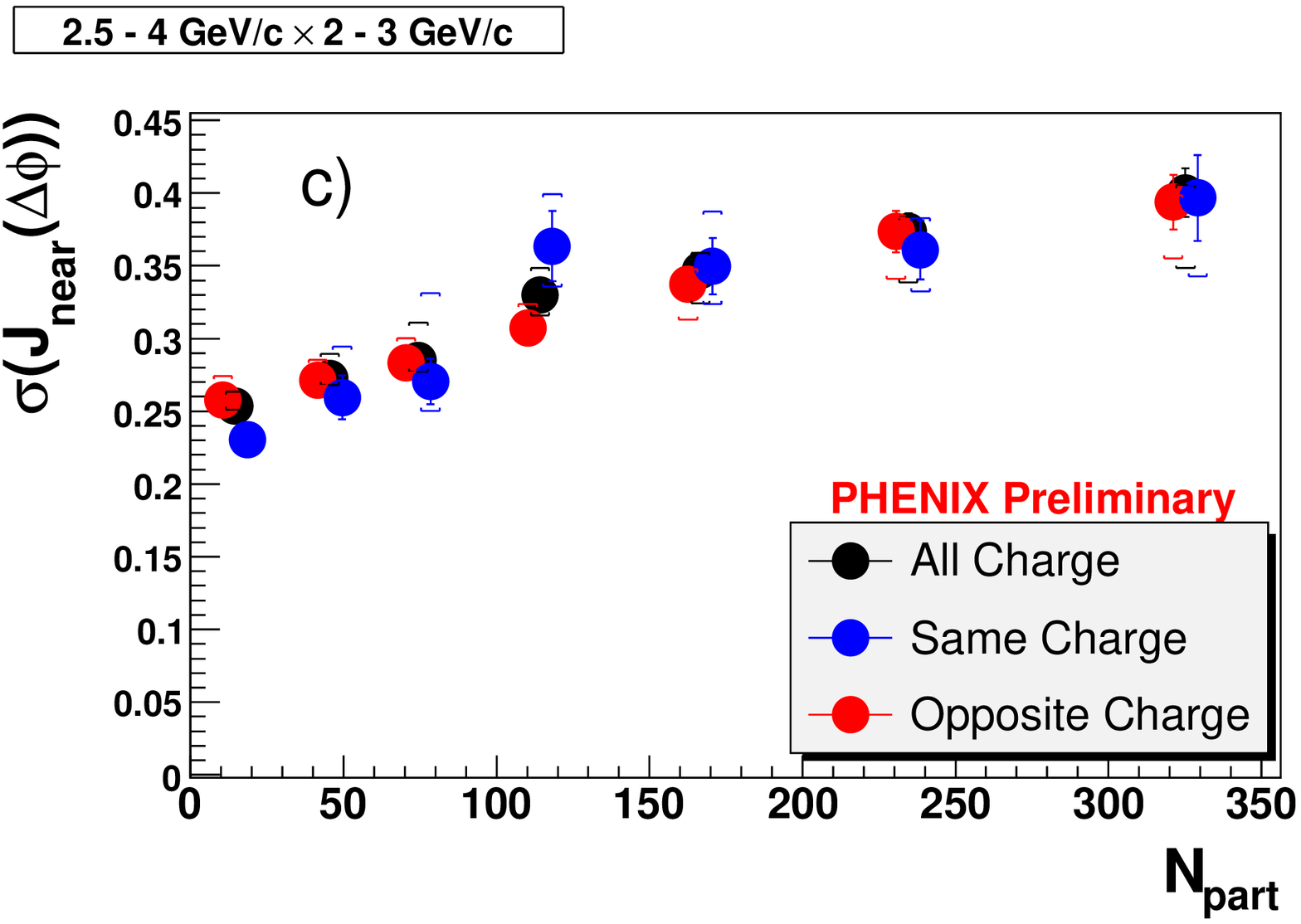,width=0.33\linewidth}
\caption{\label{fig:zyamcent} a) The correlation function for
central, mid-central and peripheral bin in Au + Au. b) The away
side ``D'' parameter as function of $N_{\rm{part}}$ for several
collision systems and energies. c) the near side width as function
of centrality for same charged pairs, opposite charged pairs and
all pairs in Au + Au.}
\end{figure}

PHENIX performed a systematic study of the jet shape and yield at
intermediate $p_T$, as shown in Fig.\ref{fig:zyamcent}. There is a
continues evolution of the split and the dip as function of
centrality. The away side split is characterized by the split
parameter $D$~\cite{nathan}, which is obtained by a double gauss
fit on the away side. $D$ seems to turn on rather quickly as a
function of centrality, and fall on a uniform curve as function of
N$_{\rm{part}}$ for different collision energies and collision
systems.

What is the nature of the away side dip? Energy loss models that
implement the jet broadening in a random walk manner can not
describe the dip or the flat jet shape~\cite{Vitev:2005at}. Models
with Cherenkov gluons~\cite{Dremin:1979yg,Koch:2005sx} or medium
dragging effect from flow~\cite{Armesto:2004vz} predict a cone or
a broadening of the away side jet, but the predicted modifications
depend strongly on momentum and are expected to disappear at large
$p_T$. Casalderrey {\it{et. al.}} \cite{Casalderrey-Solana:2004qm}
proposed a `mach cone'/`shock wave' mechanism to explain the away
side jet shape. In this model, energetic jets, which travel faster
than speed of sound ($c_s$) in the medium, excite shock waves at
an angle $\theta = \cos^{-1}\left(c_s/c\right)$. The direction of
the cone is independent of the $p_T$, but the width of the cone is
predicted to narrow for higher $p_T$.

Fig.\ref{fig:zyamcent} also indicates a sizable broadening of the
near side jet shape in central collisions. This modification is
not as dramatic as that for the away side jet, most likely due to
the surface emission bias~\cite{Drees:2003zh} in which the average
distance travelled by the near side jet is much smaller than that
for the away side jet. However, the relatively small amount of
medium that the near side jet has to go through could already lead
to some broadening. On the other hand, the baryon yield is
enhanced at intermediate $p_T$ in central
collisions~\cite{Adler:2003kg}. Since the near side jet structure
could be different between baryon trigger and meson
trigger~\cite{Adler:2004zd}, the broadening of the near side jet
width could be a consequence of the strongly modified particle
composition in central collisions.

\begin{figure}[t]
\epsfig{file=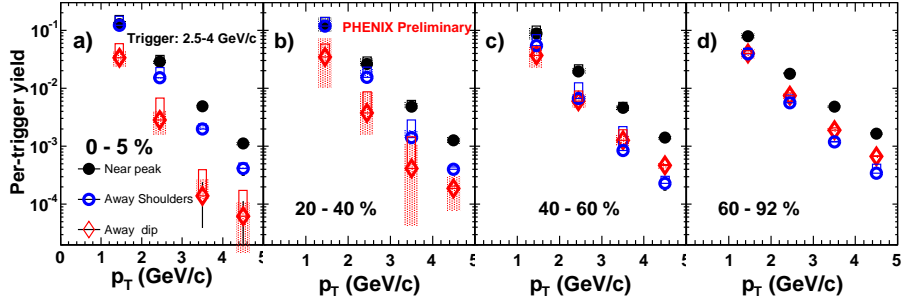,width=0.8\linewidth}
\caption{\label{fig:yield} The yield for trigger 2.5-4 GeV/$c$
plotted as function of associated hadron $p_T$ for four different
centrality bins.}
\end{figure}

To quantify the modifications of the jet shape, we study the jet
yield in three different $\Delta\phi$ regions: near side jet
region ($|\Delta\phi|<\pi/3$), the away side dip region
($|\Delta\phi-\pi|<\pi/6$), and the away side shoulder region
($|\Delta\phi-\pi\pm\pi/3|<\pi/6$). The shoulder region is
sensitive to the novel medium effects, while the dip region is
sensitive to the punch through jet contribution.
Fig.\ref{fig:yield} plots the jet yields in the three regions as
function of $p_T$ for four centralities. In 0-5\% centrality bin,
there is a large separation between the yields for the dip region
and near side jet region, persistent to large $p_T$. In more
peripheral collisions, the yield of the dip region becomes closer
or even exceeds that for the shoulder region, consistent with the
returning of the away side jet to a normal gauss shape.

\section{Dependence on the Reaction Plane.}

The study of the jet yield as function of angle w.r.p to reaction
plane is very important in the sense that it adds another
dimension in controlling the path length dependence. It also
provide additional constrains on the subtraction of the elliptic
flow background. When the trigger particles are selected in a
window centered around $\phi_s$ with a width of $\pm c$ with
respect to the reaction plane, the pair distribution up to second
order harmonics (without jet contribution) is
~\cite{Bielcikova:2003ku}:
\[
\quad \quad \frac{{dN^{\rm{pairs}} }}{{d\Delta \phi }} =
\frac{{2c}}{\pi }B(a + 2v_2^ab\cos 2\Delta \phi )
\]
\[
\begin{array}{l}
 \left\{ \begin{array}{l}
 a = 1 + 2v{_2 }^t \cos 2\phi _s \frac{{\sin 2c}}{{2c}}\left\langle {\cos 2\Psi } \right\rangle  \\
 b =  v{_2 }^t  + \cos 2\phi_s \frac{{\sin 2c}}{{2c}}\left\langle {\cos 2\Psi } \right\rangle  + v{_2 }^t \cos 4\phi _s \frac{{\sin 4c}}{{4c}}\left\langle {\cos 4\Psi } \right\rangle  \\
 \end{array} \right. \\
  \\
 \end{array}
\]

$a$ is the combinatoric background level, and it is proportional
to the number of trigger particles in the window. The correlation
function without jet contribution is:
\begin{eqnarray}
\label{eq:3}
 C(\Delta\phi) = \frac{dN^{pairs}/d\Delta \phi
}{dN^{mix}/d\Delta \phi} = \xi(1 + 2v{_2 }^ab/a\cos 2\Delta \phi)
= \xi(1 + 2v_2^av_{2,eff}^t\cos 2\Delta \phi) \
\end{eqnarray}
Where $v_{2,eff}^t = b/a$ is the effective $v_2$ of the trigger
particle in the window, and $\xi$ is the same normalization factor
as in Eq.\ref{eq:1}. $\xi$ does not depend on the trigger
direction.

\begin{figure}[t]
\epsfig{file=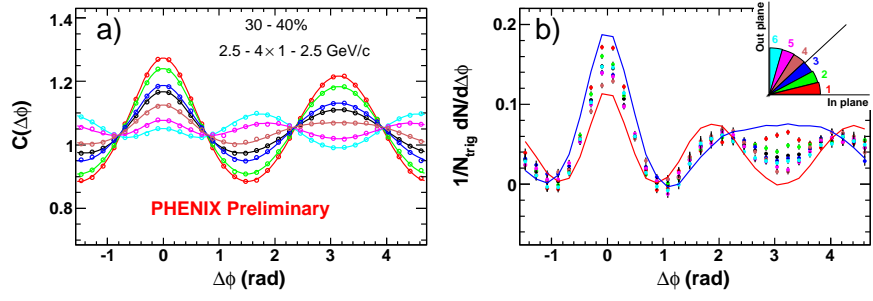,width=0.8\linewidth}
\caption{\label{fig:rp1} a) Correlation function for 6 trigger
direction bins and the trigger integrated bin (the center curve).
b) The flow subtracted per-trigger yields, the insert figure shows
the 6 trigger bins.}
\end{figure}

We divide the trigger range, [0,$\pi/2$], into 6 bins. Each bin
has a different flow background, which can be calculated from
Eq.\ref{eq:3}. The measured correlation functions for 30-40\%
centrality bin are shown Fig.\ref{fig:rp1}. Several interesting
features can be readily identified. The effective
$v_{2,\rm{eff}}^t$ changes dramatically from in plane to out of
plane direction, but the six CFs cross each other at $\pm\pi/4$
and $\pi\pm\pi/4$, where the harmonic contributions are zero. The
away side cross points are systematically higher than those at the
near side, reflecting directly the amount of jet contribution at
$\pi\pm\pi/4$. The extremes of the distributions are not at
$\pi/2$ where the flow influence is maximal, instead they are
shifted either to the left or the right due to the jet
contribution.

\begin{figure}[t]
\epsfig{file=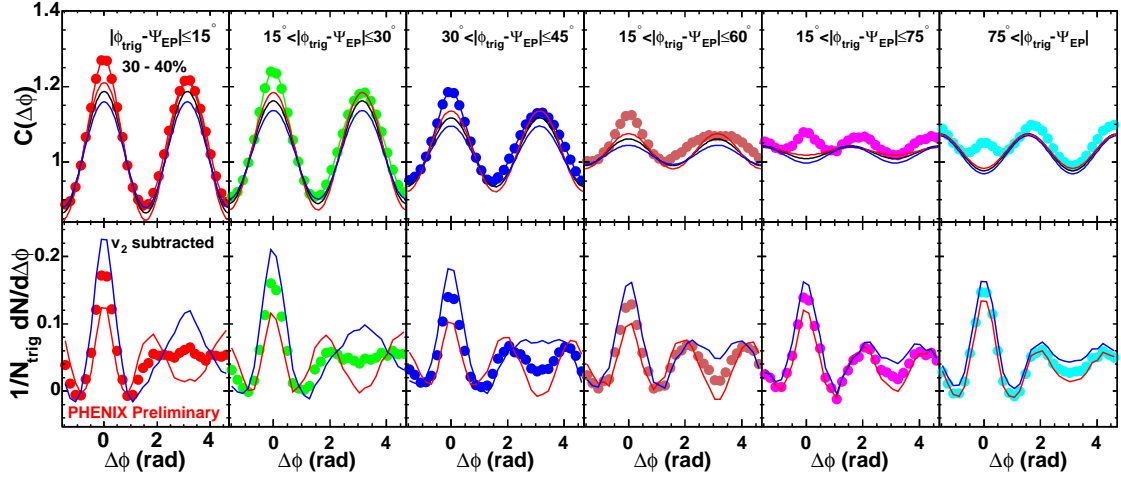,width=1\linewidth}
\caption{\label{fig:rp2} (top row) Correlation functions and
(bottom row) background subtracted per-trigger yields for the 6
trigger direction bins.}
\end{figure}

The jet yields in each trigger direction (Fig.\ref{fig:rp1}b ) are
obtained by subtracting the flow contribution. The only free
parameter, $\xi$, is fixed by the ZYAM procedure from the
integrated bin, and the flow terms in all six trigger bin are
automatically fixed (Eq.\ref{eq:3}). Fig.\ref{fig:rp2} shows the
comparison of the the measured CFs and the calculated flow
contributions for 30-40\% centrality bin. The systematic error
bands correspond to the error of the RP $v_2$, propagated
according to Eq.\ref{eq:3}. The size of the systematic errors is
largest for in plane bin and smallest for the out of plane bin. RP
dependence study helps to constrain the $v_2$ systematic when it
is not dominated by the RP resolution~\footnote{Since $v_2^a =
v_{2,\rm{raw}}^a/\left<\cos2\Psi\right>$, the error of $v_2$ from
RP resolution is independent of trigger direction.}. In
Fig.\ref{fig:rp1}b, jet shapes for different bins show some subtle
differences within the systematic error. We believe they are
mostly due to the small $v_4$ terms which were not considered in
current analysis.

\begin{figure}[ht]
\epsfig{file=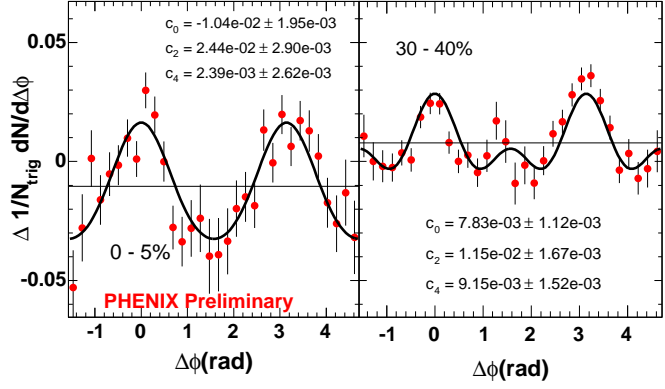,width=0.6\linewidth}
\caption{\label{fig:rp3} Difference of the in plane and out of
plane per-trigger yield for a) 0-5\% and b) 30-40\% centrality
bins.}
\end{figure}

In Fig.\ref{fig:rp3}, the difference of the per-trigger yield
between in plane direction and out of plane direction:
$(1/N_{\rm{trig}} \Delta N/\Delta\phi)_{\rm{in}} -
(1/N_{\rm{trig}} \Delta N/\Delta\phi)_{\rm{out}}$ is plotted, and
they are fitted with $c_0+c_2\cos2\Delta\phi+c_4\cos4\Delta\phi$
term. Indeed most of the variations can be accounted for by this
function. There is a small excess on the away side relative the
near side in 30-40\% centrality bin. This excess could be the hint
for the path length dependence of the away side jet modification.

\section{``Reappearance'' of the away side jets at high $p_T$}
The importance of high $p_T$ correlation is two fold. On the one
hand, high $p_T$ jets are free from complicated intermediate $p_T$
physics (for example recombination), thus can serve as a cleaner
probe of the medium. On the other hand, Studies of high $p_T$ jets
can help to disentangle normal jet fragmentation from cherenkov
gluons, shock wave or fragmentation of radiated gluons which
become dominating at intermediate or low $p_T$.

\begin{figure}[t]
\begin{minipage}{0.76\linewidth}
\begin{flushright}
\epsfig{file=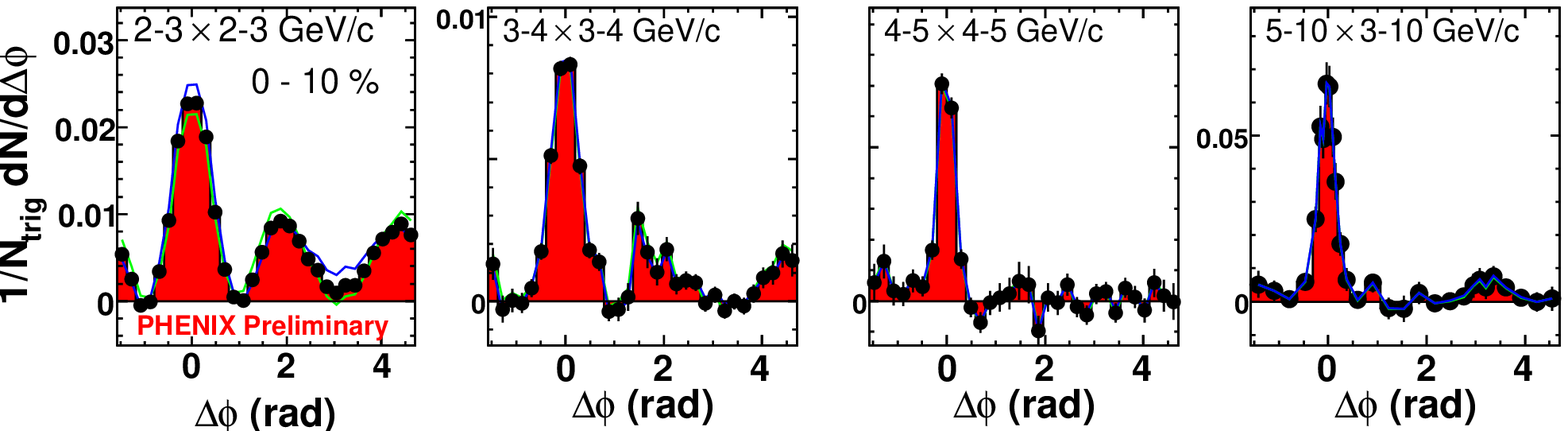,width=1\linewidth}
\end{flushright}
\end{minipage}
\begin{minipage}{0.24\linewidth}
\begin{flushleft}
\epsfig{file=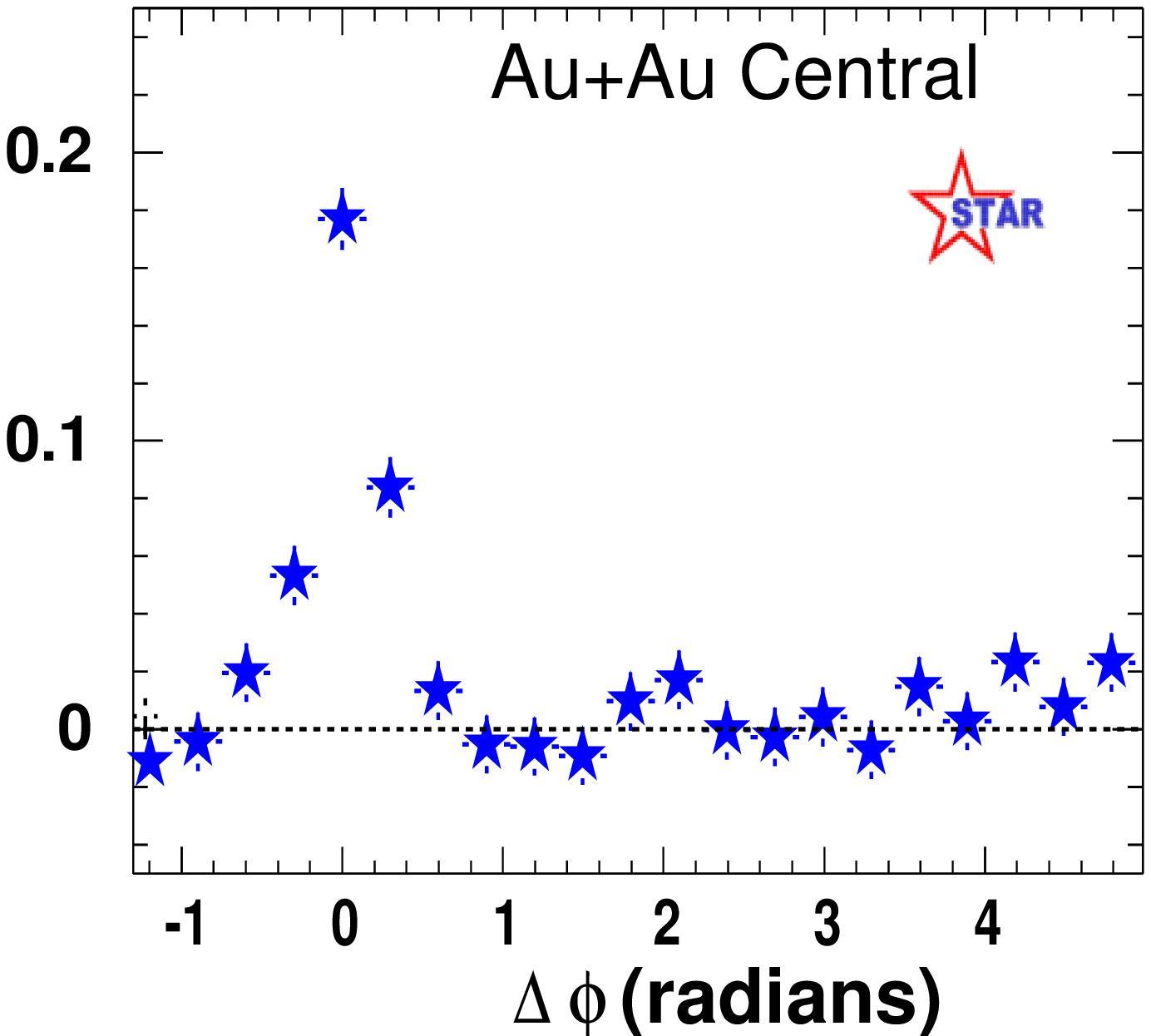,width=1\linewidth}
\end{flushleft}
\end{minipage}
\caption{\label{fig:scan1} (left four panels) : per-trigger yield
for different $p_T$ selection in 0-10\% centrality bin. (right
panel): per-trigger yield from STAR in 0-10\% centrality
bin~\cite{Adler:2002tq}.}
\end{figure}

Fig.\ref{fig:scan1} shows several CFs in successively higher $p_T$
ranges in 0-10\% central $\rm{Au + Au}$ collisions. The typical
away side cone structure persists to $p_T\approx4$ GeV/$c$, but
the edges of the cones become sharper and their magnitude drops.
In $4-5 \times 4-5$ GeV/$c$ bin, the relative flat away side shape
does not rule out the cone shape, but it's magnitude must be
significantly reduced.

As a comparison, in the most right panel of Fig.~\ref{fig:scan1},
we also show the hadron-hadron correlation from
STAR~\cite{Adler:2002tq}. The data are for 0-10\% most central Au
+ Au with $4<p_{T,\rm{trig}}<6$ GeV/$c$ and
$2<p_{T,\rm{assoc}}<p_{T,\rm{trig}}$, and are comparable to the
middle panel in Fig.~\ref{fig:scan1}. It is also almost comparable
to the highest $p_T$ point in Fig.~\ref{fig:yield}a, with the
$p_T$ selection of the trigger and associated hadrons are
swapped~\footnote{When the $p_T$ range of trigger and associated
particle are swapped, the di-jet modification factor $I_{\rm{AA}}$
are connected to each other by, $I^1_{\rm{AA}}R^1_{\rm{AA}} =
I^2_{\rm{AA}}R^2_{\rm{AA}}=\frac{\rm{Jet
pairs}_{AA}}{N_{\rm{coll}}\rm{Jet pairs}_{pp}}$, where
$R^1_{\rm{AA}}$ and $R^2_{\rm{AA}}$ are the nuclear modification
factor of the first and second particle,, respectively.
JetPairs$_{\rm{AA}}$ and JetPairs$_{\rm{pp}}$ represent the
average number of jet pairs in one A + A collision and one p + p
collision, respectively, respectively.}. All three are
qualitatively similar to each other. Interestingly, STAR's data
are consistent with zero around $\pi$, but it seems to have a
shoulder at $\pi\pm1$ as suggested by the right panel of
Fig.~\ref{fig:scan1}.

\begin{figure}[t]
\epsfig{file=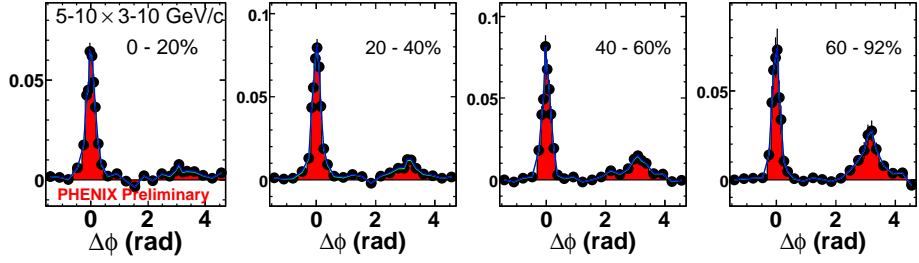,width=0.85\linewidth}
\caption{\label{fig:scan2} Centrality dependence of the
per-trigger yield at high $p_T$}
\end{figure}

\begin{figure}[t]
\epsfig{file=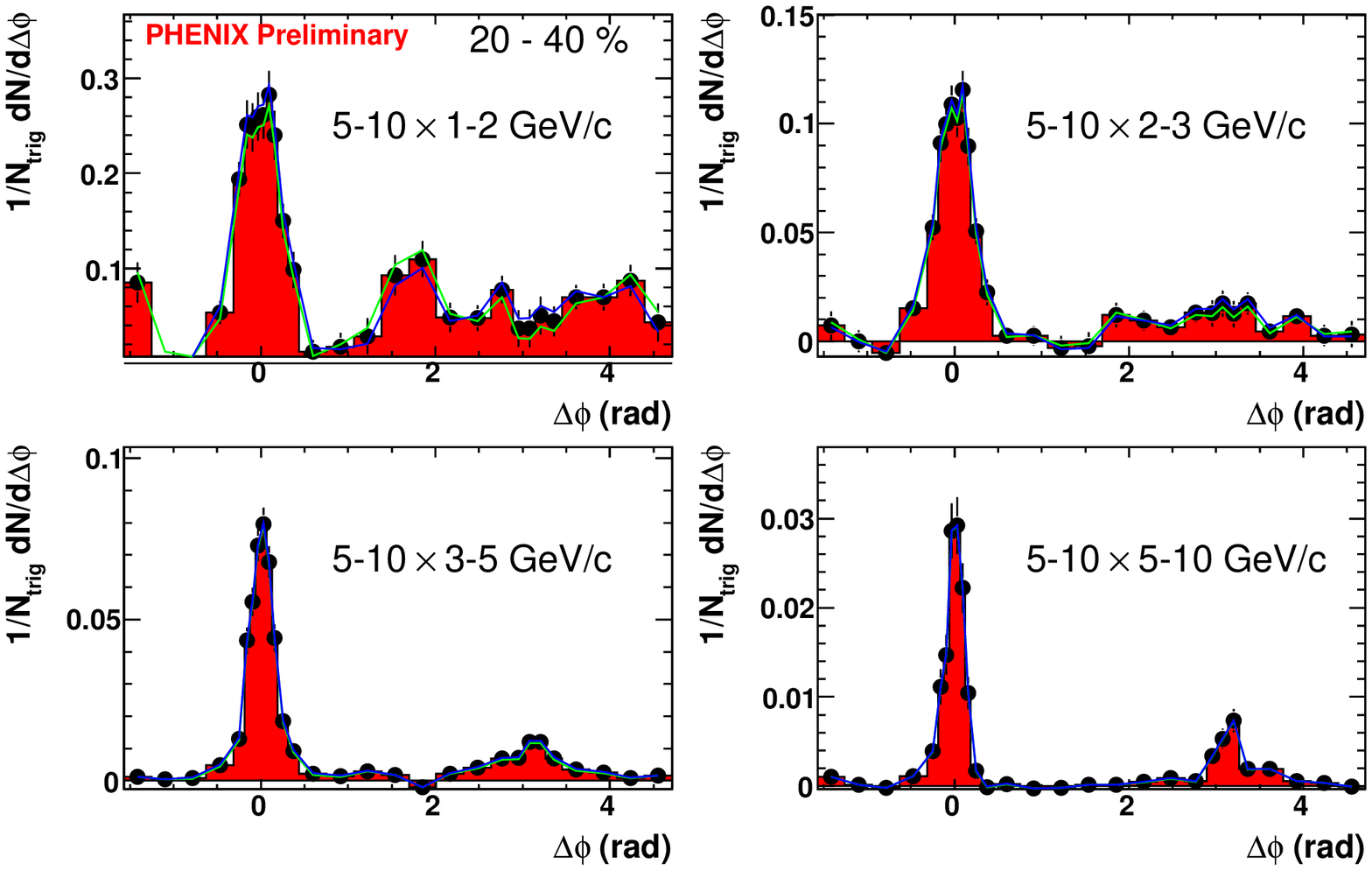,width=0.57\linewidth}
\epsfig{file=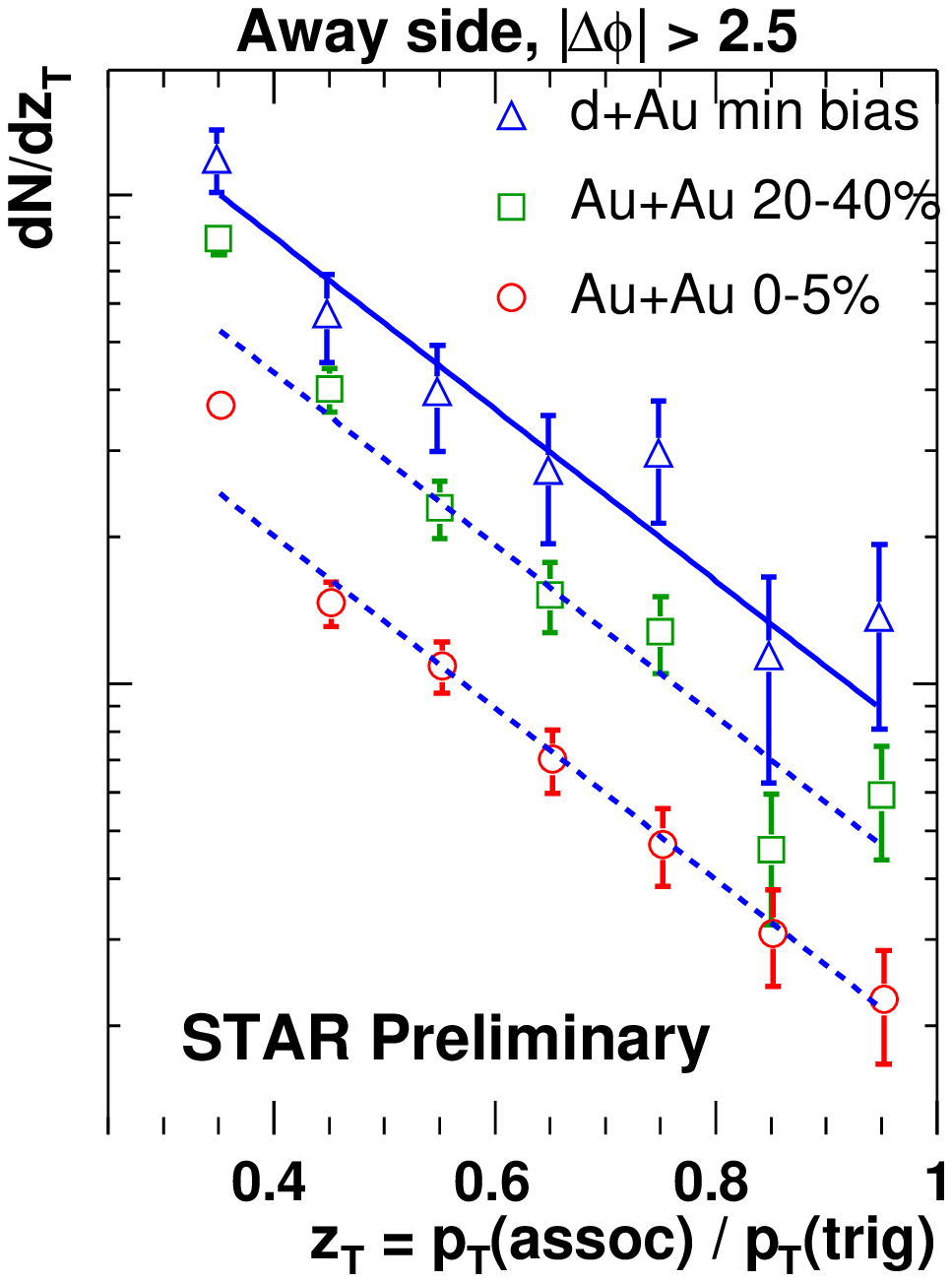,width=0.28\linewidth}
\caption{\label{fig:scan3} (left four panels): per-trigger yield
for different $p_{T,\rm{assoc}}$ in 20-40\% centrality bin when
trigger $p_T$ is fixed. (right panel) Away side jet fragmentation
from STAR~\cite{magestro}.}
\end{figure}

In the highest $p_T$ bin of Fig.\ref{fig:scan1}, a peak structure
seems to reemerge around $\pi$ on top of a flat background. To
understand the physics behind the peak structure, we plot in
Fig.\ref{fig:scan2} the centrality dependence of the CF for $5-10
\times 3-10$ GeV/$c$ selection. The away side peak exists in all
centrality bins, although its magnitude is suppressed toward
central collisions. At this point, it is hard to say whether the
widths of the away peaks are also broadened in central collisions.
On the other hand, there seems to be little change in both the
shape and magnitude of the near side jet as function of
centrality. These can be compared with recent di-jet results
measured at much larger $p_T$ ($8<p_{T,\rm{trig}}<15$ GeV/$c$ and
$6<p_{T,\rm{assoc}}$ GeV/$c$) from STAR
experiment~\cite{magestro}, where the jet width and the shape of
the fragmentation function are found to be independent of $p_T$ at
fairly large $z$ ($z>0.4-0.5$). In energy loss picture, the large
$z$ requirement biases the detected away side jets to smaller
energy loss, which biases detected jet towards surface, thus
points to the picture where both jets are emitted tangential to
the surface. If this scenario is true, we should recover the
strong medium modification at low $z$ (by decreasing
$p_{T,\rm{assoc}}$). To check this, in Fig.\ref{fig:scan3}a we
show the CF for various associated hadron $p_T$ with trigger $p_T$
fixed. The $\left<z\right>$ of the associated hadron in the four
panels are approximately 0.2, 0.4, 0.6 and 1. Clearly we see a
stronger distortion of the away side jet shape at smaller
$p_{T,\rm{assoc}}$, the yields relative to the near side are also
larger at smaller $p_{T,\rm{assoc}}$. In fact, the fragmentation
functions from STAR also suggest a significant deviation from the
uniform scaling shape at $z\lesssim0.4$ as shown in
Fig.\ref{fig:scan3}b. It is important to measure the fragmentation
function in full $z$ range in order to separate these two
competing effects.

\section{Conclusions}
Jet properties from hadron-hadron correlation have been studied as
function of $p_T$, centrality and the angle relative to the
reaction plane. Precise extraction of jet signal relies on
experimental control on the flow backgrounds, which can be
constrained by looking their reaction plane dependence. Jet shape
and yield are found to be strongly modified at intermediate and
low $p_T$. The interpretations of these modification, however, are
complicated by various competing mechanisms. By increasing the
$p_T$ for both triggering and associated hadrons, away side jet
peak reappears but it's yield is suppressed. This might be due to
the bias effect where the detected di-jets are emitted tangential
to the surface.

\end{document}